\documentclass[amsmath,amssymb,aps,prx,twocolumn]{revtex4-2}

\usepackage{graphicx}
\usepackage{dcolumn}
\usepackage{bm}

\usepackage{color}
\usepackage[english]{babel}
\usepackage[
colorlinks=true,
urlcolor=blue,
citecolor=blue,
linkcolor=blue,
hyperfootnotes=false]{hyperref}
\usepackage{hyperref}

\begin{document}

\title{Holographic Aspects of Dynamical Mean-Field Theory
}

\author{Kouichi Okunishi${}^{1,2}$}\email{okunishi@omu.ac.jp}
\author{Akihisa Koga${}^{3}$}
\affiliation{${}^1$Department of Physics, Osaka Metropolitan University, 3-3-138 Sugimoto, Osaka 558-8585, Japan}
\affiliation{${}^2$Nambu Yoichiro Institute of Theoretical and Experimental Physics (NITEP), Osaka Metropolitan University, 3-3-138 Sugimoto, Osaka 558-8585, Japan}
\affiliation{${}^3$Department of Physics, Institute of Science Tokyo, Meguro, Tokyo, 152-8551, Japan}
\date{\today}

\begin{abstract}
Dynamical mean-field theory (DMFT) is one of the most standard theoretical frameworks for addressing strongly correlated electron systems.
In this study, we explore a holographic renormalization-group (RG)-like structure inherent in DMFT, which we refer to as ``AdS/DMFT'',  by focusing on the background Bethe-lattice network behind DMFT for electrons with a semicircle density of states.
We formulate an RG transformation for the branch Green's function from the outer edge to the interior of the background Bethe-lattice network, and then find that its fixed point can be interpreted as a self-consistent solution of Green's function in DMFT.
Moreover, we clarify that the scaling dimensions for the branch Green's function and the boundary correlation functions of electrons at the outer edge of the Bethe-lattice network are characterized by the fixed-point Green's function, analogous to the behavior of a scalar field in an effective two-dimensional anti-de Sitter space.
We also perform DMFT computations for the Bethe-lattice Hubbard model, which illustrate that the scaling dimensions capture the Mott transition in the deep interior.
\end{abstract}

\maketitle

\section{introduction}

Recently, several significant advances in numerical simulation techniques, such as tensor networks for studying quantum many-body systems~\cite{JPSJ2022,Orus2019,Banuls2023}, have often been achieved in synchronization with the rapid development of quantum information technologies. 
Interestingly, such quantum information-inspired advancements also elucidate their intrinsic and nontrivial relations to various interdisciplinary research fields in fundamental physics.
For example, the Ryu-Takayanagi formula for holographic entanglement entropy~\cite{RT_PRL2006,RT_JHEP2006}, which originates from string theory, extracts universal properties of quantum entanglement, such as the area law of entanglement entropy~\cite{Eisert_RMP2010}, and provides a guiding principle for designing practical algorithms of the tensor network renormalization group (RG) for critical systems~\cite{MERA2007,TNR2015,Swingle2012}. 
Moreover, several intriguing ideas, such as quantum chaos and its connection to quantum information scrambling in black holes~\cite{SY1993,Sachdev2015},  are shared among condensed matter, quantum information, and quantum gravity research fields, mediated by anti-de Sitter/Conformal Field Theory (AdS/CFT)~\cite{Maldacena1998,Aharony2000,Witten1998,Gubser1998}.
The above holographic nature underlying various quantum many-body systems motivates us to explore further theoretical and numerical methods where a holographic mechanism plays a pivotal role in their background.

In this paper, we shed light on the holographic RG-like structure inherent in dynamical mean-field theory (DMFT)~\cite{Metzner_1989,Muller_1989,Pruschke_1995,DMFT_RMP1996}, which we refer to as ``AdS/DMFT''.
The DMFT is one of the most standard and widely used theoretical frameworks for addressing strongly correlated electron systems, a central issue in modern condensed matter physics.
\footnote{In the context of DMFT, the term ``dynamical'' refers to the self-consistent treatment of the frequency-dependent local self-energy arising from electron-electron interactions based on the imaginary-time path-integral formalism. It does not imply any dynamical coupling between the electron degrees of freedom and the geometric (gravitational) structure.}
DMFT is formulated on the basis of the local Green's function in an effective medium, and the incorporation of numerical techniques developed for the quantum impurity problem~\cite{WilsonRG,HirschFye1986,LoopQMC_imp2011,Hoffstetter2000,Ganahl2014,Bauernfeind2017} facilitates quantitative and reliable treatments of quantum many-body effects in the bulk.
Indeed, extensive applications to various Hubbard-type models demonstrated several characteristic behaviors induced by strong electron interactions, such as the Mott transitions~\cite{Zhang_1993,Bulla1999,Koga2004,Werner2007},
magnetically ordered states~\cite{Obermeier_1997,Momoi_1998,Chitra_1999,Zitzler_2002,Yanatori_Koga_2016,Kamogawa_2019,Fujii_2025},
and superconducting states~\cite{Georges_1993,Capone_2002,Koga_2015,Hoshino_Werner_2017,Ishigaki_2018}.

\begin{figure}[tb]
\includegraphics[width=6cm]{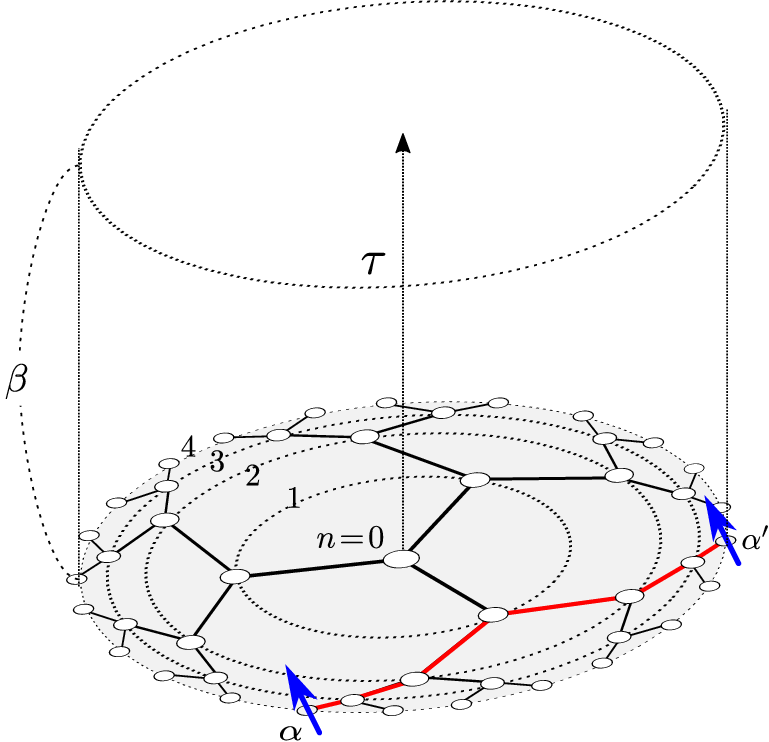}~~
\caption{The Bethe lattice with $q=3$ and $N=4$ with the imaginary time axis, where the index $n$ for dotted circles represents the generation from the center node and $\tau$ denotes the imaginary time.
Note that the antiperiodic boundary condition is assumed in the $\tau$ direction, where $\beta$ denotes an inverse temperature.
The distance between $\alpha$ and $\alpha'$ in the circumference direction is defined by counting the number of nodes along the outer edge, while the red line represents the shortest network path connecting the edge spins.
}
\label{tag_fig1}
\end{figure}

A key point in discussing AdS/DMFT is that many-body Green's functions are effectively factorized into local Green's functions in the limit of infinite spatial dimension. 
This factorization also implies that DMFT for the semicircle density of states, or equivalently, the Bethe-lattice-type network, captures intrinsic correlation effects in the strongly correlated electron system.
Then, an essential feature of the Bethe-lattice network is that a loop-free network structure and translational invariance coexist at the level of an infinite-size tree network, and the number of nodes increases exponentially toward the outer edge (see Fig.~\ref{tag_fig1}),  which allows us to interpret the Bethe-lattice Ising model as a toy model in the side of Euclidean AdS$_2$~\cite{Oku2023}, consistent with a Burhat-Tits tree in the context of $p$-adic AdS/CFT~\cite{Gubser2017,Heydeman2016,Bhattacharyya2017,Bentsen2019}.

Inspired by the observations above, we set up an RG transformation by taking a partial sum from the outer edge to the interior of a correlated-electron system on a background Bethe-lattice network.
Assuming the effective factorization of Green's function for a descendant branch in the network as in DMFT, we derive an RG relation for the branch Green's function on the Bethe lattice, which also corresponds to the iterative update of an effective chemical potential in a DMFT calculation.
We then find that the bulk fixed point equation of the RG transformation corresponds to the self-consistent relation in DMFT.
Moreover, we can calculate the correlation function of the electron-density operators, which involves the response of the system to a perturbation of the chemical potential at the outer edge node.
By appropriately smearing out lattice nodes in the network, moreover, one can also introduce an effective coordinate corresponding to Poincar\'e's upper half~\cite{Oku2023}.
We then extract the scaling dimensions associated with the outer-edge chemical potential analogous to the behavior of a scalar field in an effective AdS$_2$, which can be also related to the convergence of the DMFT iteration.
These suggest that the DMFT for the correlated-electron system can also serve as a prototypical lattice network model that incorporates quantum effects via the imaginary-time path integral, providing a complementary viewpoint for quantum effects on the AdS$_2$.

This paper is organized as follows.
In the next section, we reformulate DMFT for the Bethe-lattice model based on the standard path-integral representation~\cite{DMFT_RMP1996}.
The tree network nature\cite{Hikihara2023,Hikihara2025,TTN2006} of the Bethe lattice network enables us to construct an effective environment in DMFT through a recursive RG transformation for the Green's function of the descendant branches.
We then derive the fixed point of the RG transformation for the branch Green's function, which eventually determines the self-consistent solution of the local Green's function in DMFT.
We also provide the linearized RG equation around the fixed point, as well as the RG transformation for an electron operator at the outer-edge boundary of the Bethe lattice network.
In Sec. 3, we further extract the scaling dimensions of the correlation functions for outer-edge electrons, incorporating the effective Poinca\`re coordinate introduced for the Bethe-lattice network.
In Sec.~4, we demonstrate the $U$ dependence of the scaling dimension the Bethe-lattice Hubbard model through a DMFT computation with the semicircle density of state, where $U$ denotes the electron interaction\cite{Bulla1999,Werner2007}.
In Sec.~5, we summarize the result and its implications, and then discuss a further extension of the holographic RG structure associated with DMFT. 
In Appendix A, we present the analytic expression of the RG for the free-electron case.

\section{ A holographic-RG structure in DMFT}

We reformulate DMFT inspired by the holographic RG point of view.
Using the tree-network nature of the Bethe lattice model, we set up a recursive relation for branch Green's functions by sequentially taking partial sums for electrons from the outer edge boundary nodes. 
The details of this formulation are presented in the following.

\subsection{formulation}

Let us start with the path-integral representation of the partition function for the Bethe lattice Hubbard model.
The Hamiltonian in operator form is 
\begin{align}
\mathcal{H} =& -t\sum_{\langle i,j\rangle\sigma}[ c^\dagger_{i\sigma} c_{j\sigma} + c^\dagger_{j\sigma} c_{i\sigma}] \nonumber \\
 &  - \mu \sum_{i \sigma} c^\dagger_{i\sigma} c_{i\sigma} + U\sum_i c^\dagger_{i\uparrow} c_{i\uparrow} c^\dagger_{i\downarrow} c_{i\downarrow} \label{H}
\end{align}
where $t$, $\mu$ and $U$ denote a hopping parameter, a chemical potential, and the Hubbard interaction, respectively.
Also, $c^\dagger_{i\sigma}$ ($c_{i\sigma}$) is a creation (annihilation) operator of an electron with spin $\sigma$ at node $i$, and $\langle i,j\rangle$ is the sum of nearest-neighbor pairs on the Bethe lattice.

\begin{figure}[bt]
\includegraphics[width=5cm]{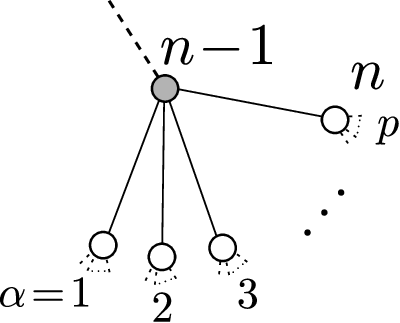}~~
\caption{The network connectivity for a parent node [($n-1$)th generation]} and its descendant branch [$n$th generation] indicated by $\alpha=1, 2,\cdots, p$. We have omitted the branch index for the parent node. 
\label{tag_fig2}
\end{figure}

As shown in Fig.~\ref{tag_fig1},  we introduce the coordination number $q$, the branching number $p\equiv q-1$, and the generation index $n=0, \cdots N$, which are useful for efficiently describing the Bethe lattice model.
Here, note that $n=0$ and $N$ correspond to the center and the outer edge of the Bethe lattice, respectively.
In DMFT $q, N \gg 1$ is assumed. 
We also introduce Grassmann variables $\bar{c}_{n \sigma\alpha}$ and ${c}_{n \sigma\alpha}$ respectively corresponding to electron creation and annihilation operators $c^\dagger_{n\sigma\alpha}(\tau)$ and $c^{}_{n\sigma\alpha}(\tau)$ at a $n$th generation node, where $\alpha=1, \cdots p$ indicates a descendant branch index connected with a node at a $(n-1)$th generation node (see Fig.~\ref{tag_fig2}).
In the following, the imaginary time $\tau$ for the Grassmann variables is omitted for simplicity, except for the case where $\tau$ is explicitly specified. 
Then, we can write the partition function $Z$ as

\begin{align}
    Z = \int \ D\bar{c}_0 Dc_0   e^{-S_0} \left[ g_N(\bar{c}_0,c_0) \right]^q
    \label{eq_def_Z}
\end{align}
by introducing the recursive relation
\begin{align}
    g_{N-n+1}(&\bar{c}_{n-1}, c_{n-1}) \nonumber \\
    &=  \int D\bar{c}_{n} Dc_{n} e^{ -\Delta S_{n-1,n} -S_{n} } g_{N-n}(\bar{c}_{n}, c_{n}) \, ,
\label{eq_def_grecursion}
\end{align}
for $ 0<  n  \le N$ with the ``boundary condition" 
\begin{align}
    g_{0}(\bar{c}_N, c_N) = e^{ -S_N }  \, .
\end{align}
Here, $Dc_n D\bar{c}_n$ represents the path-integral measure for electrons of the $n$th generation, for which we omit the indices of $\sigma$ and $\alpha$ for simplicity.
In Eqs. (\ref{eq_def_Z}) and (\ref{eq_def_grecursion}), the actions are defined by 
\begin{align}
& S_n =  \int_0^\beta d\tau \sum_{\sigma\alpha}\bar{c}_{n\sigma\alpha} (\partial_\tau - \mu) c_{n\sigma\alpha} + U n_{n\uparrow\alpha}n_{n\downarrow\alpha} \, , \label{eq_onsiteS}\\
&\Delta S_{n-1,n}= \int_0^\beta d\tau  \sum_{\sigma\alpha} -t (\bar{c}_{n-1\sigma}c_{n\alpha\sigma}+ \bar{c}_{n\sigma\alpha} c_{n-1\sigma}) \, ,
\label{eq_hoppingS}
\end{align}
where the branch index for the parent node [($n-1$)th generation] was suppressed as in Fig.~\ref{tag_fig2}.
Note that Eqs. (\ref{eq_onsiteS}) and (\ref{eq_hoppingS}) respectively represent the action for on-site terms and the electron hopping term between the two adjacent generations.

Using the recursive relation above, we consider the coarse graining of electron degrees of freedom from the outer edge to the interior.
The first step is to calculate
\begin{align}
    g_{1}(\bar{c}_{N-1}, c_{N-1}) = \int D\bar{c}_{N} Dc_{N} e^{ -\Delta S_{N-1,N} -S_{N}  }
\end{align}
where
\begin{align}
    S_N &= \int_0^\beta d\tau \sum_{\sigma\alpha}\left[   \bar{c}_{N\sigma\alpha}(\partial_\tau - \mu) c_{N\sigma\alpha} + U  n_{N\alpha \uparrow } n_{N\alpha \downarrow}  \right]     \,.
\end{align}
represents the action for the single-node Hubbard model on the outer edge.

As in DMFT,  we construct an effective medium at the parent node in which the correlation effects at the $N$th generation are renormalized.
Here, we define 
\begin{align}
    \mathcal{H}' (\tau) &= -t\sum_{\sigma\alpha}(\bar{c}_{N-1\sigma} c_{N\sigma\alpha} + \bar{c}_{N\sigma\alpha} c_{N-1\sigma} ) \, .
\end{align}
for later convenience.
Expanding the term $\Delta S_{N-1,N}$, we then have
\begin{align}
    g_1(\bar{c}_{N-1}, c_{N-1})&= 
     Z_0 \left( 1 - \int d\tau \langle  \mathcal{H}'(\tau) \rangle_0  \right.\nonumber \\
     &\left. + \frac{1}{2} \int d\tau_1 d\tau_2 \langle \mathcal{H}'(\tau_1) \mathcal{H}'(\tau_2\rangle_0 \cdots \right) \, ,
\label{eq_g1perturbation}
\end{align}
where   
\begin{align}
\langle \cdots \rangle_0 = \frac{1}{Z_0 } \int D\bar{c}_{N} Dc_{N} \cdots e^{ -S_N } 
\label{eq_branchaverage}
\end{align}
with $Z_0 =  \int D\bar{c}_{N} Dc_{N} e^{ -S_N }$.\footnote{The suffix for $g$, $Z$ and $\langle \cdots \rangle$ represents the number of RG transformations, which is equivalent to the number of nodes from the outer edge, rather than the generation index.}
Here, note that Eq. (\ref{eq_branchaverage}) is the average within the descendant branch. 
Defining the branch Green's function as
\begin{align}
    G_{N\sigma\alpha} ( \tau-\tau' ) = -\langle T_\tau  c_{N\sigma\alpha}(\tau) {c}^\dagger_{N\sigma\alpha}(\tau') \rangle_0
\end{align}
with $T_\tau$ being the imaginary time ordering, 
we then obtain
\begin{align}
    &\langle  \mathcal{H}'(\tau_1)  \mathcal{H}'(\tau_2) \rangle_0 = \nonumber \\
     & -2 t^2 \sum_\sigma \bar{c}_{N-1\sigma}(\tau_1) c_{N-1\sigma}(\tau_2) \sum_{\alpha} G_{N \sigma\alpha}(\tau_1-\tau_2) \, .
\end{align}

In Eq. (\ref{eq_g1perturbation}), the many-body Green's function bridging different branches is already factorized, since electrons in distinct $\alpha$ branches are independent.
In contrast, the electrons within the same branch still contain a nontrivial many-body effect, which prevents a naive factorization of the many-body Green's function.
For $p \gg 1$, which is the case of DMFT, however, the weight of the connected Green's function becomes relatively negligible.
Although this situation is justified in the $p\to \infty$ limit, we may assume the free fermionic factorization of the many-body Green's function as an approximation for a finite $p$ below.
Here, let $k$ be an integer indicating the ($2k$)th order term of Eq. (\ref{eq_g1perturbation}).
Then, we have $k!$ number of equivalent permutations for ${}_{2k}C_k$ possible pairs of $\bar{c}$ and $c$ in, which yield
\begin{widetext}
\begin{align}
& \frac{1}{(2k)!} \int d\tau_1\cdots d\tau_{2k}  \langle   \mathcal{H}'(\tau_1) \cdots \mathcal{H}'(\tau_{2k}) \rangle_0  \nonumber \\
& \simeq (-)^k\frac{t^{2k} }{k!}\sum_{\sigma_1 \cdots \sigma_k} \int  d\tau_1 \cdots d\tau_k 
\bar{c}_{N-1\sigma_1}(\tau_1) c_{N-1\sigma_1}(\tau_2) \cdots \bar{c}_{N-1\sigma_k}(\tau_{2k-1}) \bar{c}_{N-1\sigma_n}(\tau_{2k}) 
\nonumber \\
& \times  
\sum_{\alpha_1\cdots\alpha_k } G_{N\sigma_1\alpha_1}(\tau_1-\tau_2) \cdots G_{N\sigma_k\alpha_k}(\tau_{2k-1}-\tau_{2k}) \, . 
\label{eq_2n_expansion_U}
\end{align}
Collecting all terms, then, we have 
\begin{align}
    &g_1(c_{N-1,\sigma},\bar{c}_{N-1\sigma}) 
    = Z_0 e^{-t^2 \sum_\sigma \int d\tau_1d\tau_2\bar{c}_{N-1\sigma}(\tau_1)c_{N-1\sigma}(\tau_2) \sum_\alpha G_{N\sigma\alpha}(\tau_1-\tau_2) } \, ,
    \label{eq_effective_N-1}
\end{align}
which implies that the electron correlation in the descendant branches can be lifted to the parent generation at the level of the effective medium. 
If all the descendant branches are equivalent, we can rewrite
\begin{align}
    g_1 (c_{N-1,\sigma},\bar{c}_{N-1\sigma}) = Z_0 e^{- p t^2 \sum_\sigma \int d\tau_1d\tau_2\bar{c}_{N-1\sigma}(\tau_1)c_{N-1\sigma}(\tau_2) G_{N\sigma}(\tau_1-\tau_2) } \, ,
\end{align}
where we have omitted the branch index $\alpha$.

Let us recover a branch index $\alpha'$ for electrons at the parent node of the ($N-1$)th generation explicitly.
Taking account of all descendant branches in the ($N-1$)th generation, then, the recursive relation becomes
\begin{align}
    g_{2}(\bar{c}_{N-2},c_{N-2}) 
    &=  \int D\bar{c}_{N-1} Dc_{N-1} e^{ -\Delta S_{N-2,N-1} -S_{N-1\alpha'} } \prod_{\alpha'} g_{1}(\bar{c}_{N-1\sigma\alpha'},c_{N-1\sigma\alpha'}) \nonumber \\
    &=  (Z_0)^p \int D\bar{c}_{N-1} Dc_{N-1} e^{ -\Delta S_{N-2,N-1}  } \nonumber \\
    &e^{-\int d\tau_1d\tau_2 \sum_{\sigma\alpha'} \bar{c}_{N-1\sigma\alpha'}(\tau_1)\left[  (\partial_\tau - \mu)\delta(\tau_1-\tau_2) + pt^2 G_{N\sigma}(\tau_1-\tau_2)  )\right]c_{N-1\sigma\alpha'}(\tau_2) 
    +  U \sum_{\alpha'} n_{N-1\uparrow\alpha'}n_{N-1\downarrow\alpha'}\delta(\tau_1 -\tau_2)   }\,,
\end{align}
in which $
     - pt^2 G_{N\sigma}(\tau_1-\tau_2)
$
is regarded as an effective self-energy for electrons in the $(N-1)$th generation.

The above recursive relation can be straightforwardly generalize to $\bar{n}(=1,\cdots, N)$, which denotes the number of RG steps counted from the outer edge node.
Here, note that the number of RG steps and the corresponding generation index $n$ satisfy $n+\bar{n} = N$.
Given $G_{N-\bar{n}+1}$, we have the effective action after $\bar{n}$ times of the RG transformation
\begin{align}
&    S_{N-\bar{n}}^{\mathrm{eff} } = -\int d\tau_1 d\tau_2 \sum_{\sigma\alpha'}
\bar{c}_{N-\bar{n}\sigma\alpha'}(\tau_1)\mathcal{G}^{-1}_{N-\bar{n}\sigma}(\tau_1-\tau_2)c_{N-\bar{n}\sigma\alpha'}(\tau_2) +\sum_{\alpha'} U n_{N-\bar{n}\uparrow\alpha'} n_{N-\bar{n}\downarrow\alpha'}\delta(\tau_1-\tau_2)\, ,
\label{eq_seffN-1_U2}\\
&\mathcal{G}^{-1}_{N-\bar{n}\sigma}(\tau_1-\tau_2) =  -(\partial_\tau - \mu)\delta(\tau_1-\tau_2) -pt^2 G_{N-\bar{n}+1\sigma}(\tau_1-\tau_2) \, .
\label{eq_greenN-1_U2}
\end{align}
Here, we have omitted the branch index $\alpha$ for $G_{N-\bar{n}\sigma}$ by assuming that all the descendant branches at $N-\bar{n}+1$ are equivalent.
Using the above effective medium action, we can then calculate the branch Green's function at $N-\bar{n}$,
\begin{align}
    G_{N-\bar{n}\sigma\alpha}(\tau_1 - \tau_2) & = - \langle c_{N-\bar{n}\sigma\alpha}\bar{c}_{N-\bar{n}\sigma\alpha}\rangle_{\bar{n}}
    \label{eq_G_recursive_U}
\end{align}
with
\begin{align}
    \langle \cdots \rangle_{\bar{n}} &\equiv \frac{1}{Z_{n}} \int D \bar{c}_{N-\bar{n}} Dc_{N-\bar{n}}  \cdots e^{ - S^{\mathrm{eff}}_{N-\bar{n}} }\, \\
            Z_{\bar{n}} &\equiv \int D \bar{c}_{N-\bar{n}} Dc_{N-\bar{n}}  e^{ - S^{\mathrm{eff}}_{N-\bar{n}} }\, . 
\end{align}
Using these notations, we obtain 
\begin{align}
        g_{\bar{n}+1}(\bar{c}_{N-\bar{n}-1}, c_{N-\bar{n}-1}) 
    &=(Z_0)^{p^{\bar{n}}}\cdots (Z_{\bar{n}-1})^p Z_{\bar{n}} e^{-pt^2 \sum_\sigma \int d\tau_1 d\tau_2 \bar{c}_{N-\bar{n}-1}(\tau_1)c_{N-\bar{n}-1}(\tau_2)  G_{N-\bar{n}}(\tau_1 - \tau_2) }\, ,
\end{align}
which defines the effective action for the $(N-\bar{n}-1)$th generation with replacing $\bar{n}\to \bar{n}+1$ in Eq. (\ref{eq_seffN-1_U2}).
We thus obtain a closed form of the recursive relation with respect to the generation index, providing a holographic RG-like structure on the Bethe lattice, where the descendant branches exponentially bunch up from the outer edge to the interior.
Finally at the center node of $\bar{n}=N$, we retract the partition function for the total Bethe lattice network as 
\begin{align}
        Z &= \int \ D\bar{c}_0 Dc_0   e^{-S_0} \left[ g_N(\bar{c}_0, c_0) \right]^q \nonumber \\
        & = (Z_{0})^{p^{N-1}q}\cdots (Z_{N-2})^{pq}(Z_{N-1})^{q}\int \ D\bar{c}_0 Dc_0   e^{-\int  d\tau_1 d\tau_2 \sum_{\sigma}
\bar{c}_{0\sigma}(\tau_1)\mathcal{G}^{-1}_{0\sigma}(\tau_1-\tau_2)c_{0\sigma}(\tau_2) + U n_{0\uparrow} n_{0\downarrow}\delta(\tau_1-\tau_2) }
\label{eq_bulk_Z}
\end{align}
\end{widetext}
from which one can extract bulk properties of the total system.
Here, note that for sufficiently large $N$, RG iterations for the branch Green's functions converge to a self-consistent solution. 
Then, if $p=(q-1)\gg 1$,  deviation between the branch Green's function and the bulk Green's function defined by Eq. (\ref{eq_bulk_Z}) becomes relatively negligible, which is the case of DMFT.

In the above set of recursion relations, the most difficult part is to calculate Eq. (\ref{eq_G_recursive_U}), which corresponds to solving a quantum impurity problem in an effective medium defined by Eq. (\ref{eq_greenN-1_U2}). 
For this problem, several numerical solvers have been developed, such as quantum Monte Carlo simulation~\cite{HirschFye1986,Werner_2006,LoopQMC_imp2011}, 
exact diagonalization~\cite{Caffarel_1994}, numerical renormalization group~\cite{Bulla1999}, and 
tensor networks~\cite{Hoffstetter2000, Ganahl2014,Bauernfeind2017}, as established in the long history of DMFT.

\subsection{Fourier representation}

As in DMFT, the Fourier representation of the brach Green's function with respect to the imaginary time $\tau$ is more useful in the RG formulation above.
Assuming no branch dependence, we write the Fourier transform for the branch Green's function at $(N-\bar{n}+1)$th generation as
\begin{align}
    G_{N-\bar{n}+1\sigma}(i\omega_l) & = \int_0^\beta d\tau e^{i\omega_l \tau} G_{N-\bar{n} +1\sigma}(\tau) \, ,
\end{align}
where $\omega_l= \frac{\pi (2l +1)}{\beta}$ with $l=0, \pm 1, \pm2 \cdots $ is the Matsubara frequency.
We then have the effective action for the ($N-\bar{n})$th generation
\begin{align}
& S_{N-\bar{n}}^{\mathrm{eff} }  = \ \sum_{\sigma} \sum_{l}     {\bar{c}}_{N-\bar{n}\sigma,l}    [\mathcal{G}_{N-\bar{n}\sigma}(i\omega_l) ]^{-1}  {c}_{N-\bar{n}\sigma,l}   \nonumber \\
    &  + \frac{U}{\beta}\sum_{jklm} {\bar{c}}_{N-\bar{n}\uparrow,j}  {\bar{c}}_{N-\bar{n}\downarrow ,l} {c}_{N-\bar{n}\downarrow,k} {c}_{N-\bar{n}\uparrow,m} \delta_{j+k-l-m,0} \, 
    \label{eq_omega_effaction}
\end{align}
with
\begin{align}
     [\mathcal{G}_{N-\bar{n}\sigma}(i\omega_l) ]^{-1}    = i\omega_l + \mu   - pt^2 G_{N-\bar{n}+1\sigma}(i\omega_l) .
     \label{eq_k-effectivemedium}
\end{align}
Here, the Fourier transform of electron variables is defined as $c_{N-\bar{n}\sigma,l} = \frac{1}{\sqrt{\beta}} \int_0^\beta d\tau e^{i\omega_l \tau} c_{N-\bar{n}\sigma}(\tau)$ and $\bar{c}_{N-\bar{n}\sigma,l} = \frac{1}{\sqrt{\beta}} \int_0^\beta d\tau e^{-i\omega_l \tau} \bar{c}_{N-\bar{n}\sigma}(\tau)$ with the antiperiodic boundary condition in the $\tau$ direction.
For $S_{N-\bar{n}}^{\mathrm{eff} }$, one can then calculate 
\begin{align}
 G_{N-\bar{n}\sigma}(i\omega_l) = \langle \bar{c}_{N-\bar{n}\sigma, l }c_{N-\bar{n}\sigma,l} \rangle_{\bar{n}}
 \label{eq_bethe_Gimpurity}
\end{align}
which completes the recursive RG relation in the $\omega$ space.

Here, we discuss the physical interpretation of Eq. (\ref{eq_k-effectivemedium}).
By introducing an effective chemical potential with 
\begin{align}
    \tilde{\mu}_{N-\bar{n}}(i \omega_l) \equiv \mu - pt^2 G_{N-\bar{n}+1\sigma}(i\omega_l) \, ,
\end{align}
we can regard Eq. (\ref{eq_k-effectivemedium}) as an RG transformation for the $\omega$- (or equivalently $\tau$-) dependent chemical potential.
Then, Eq.  (\ref{eq_k-effectivemedium}) describes how the chemical potential applied at the outer edge node is renormalized toward the interior: $\tilde{\mu}_N \to \tilde{\mu}_{N-1} \to \tilde{\mu}_{N-2} \to \cdots$.
In other words, the iterative update of ${G}(i\omega_l)$ plays the same role as the RG relation of the effective chemical potential, as in the case of the magnetic field for the Bethe-lattice Ising model.

\subsection{Fixed point and linearized RG equation} 
\label{subsec_3-C}

On the basis of the Fourier representation above, we analyze the fixed point of the recursive relation, Eq. (\ref{eq_k-effectivemedium}) and Eq. (\ref{eq_bethe_Gimpurity}).
For later convenience, we formally write the functional form of the branch Green's function $G_{N-\bar{n}}$, which is obtained by solving the impurity problem under $\mathcal{G}_{N-\bar{n}}$, as 
\begin{align}
    G_{N-\bar{n} \sigma}(z) = F[\mathcal{G}_{N-\bar{n}\sigma}(z)]\, ,
    \label{eq_def_Functional}
\end{align}
where we have introduced a complex variable $z$ instead of $i \omega_l$.
Then, we can determine the fixed point of the recursive relation as a solution of
\begin{align}
     [\mathcal{G}^*_{\sigma}(z) ]^{-1}   & = z + \mu   - pt^2 G^*_{\sigma}(z),\label{FPRR}\\
     G^*_{\sigma}(z) &=F[\mathcal{G}^*_{\sigma}(z)], 
     \label{eq_solution_for_RG}
\end{align}
which corresponds to the self-consistent equation for the deep interior of the Bethe lattice network.
Note that the solution of Eq. (\ref{eq_solution_for_RG}) for the free electron with $p\gg 1$ yields the semicircle density of states. (See Appendix \ref{appendix_A})

We next consider a perturbation around the fixed point,
\begin{align}
    G_{N-\bar{n}+1\sigma}(z) =  G^*_{\sigma}(z) + \delta G_{N-\bar{n}+1\sigma}(z) \, .
\end{align}
Linearizing Eq. (\ref{eq_k-effectivemedium}) with respect to $\delta G$, we obtain
\begin{align}
    \delta \mathcal{G}_{N-\bar{n}\sigma}(z) &= \frac{pt^2}{(i\omega_l + \mu   - pt^2 G^*_{\sigma}(z) )^2} \delta G_{N-\bar{n}+1\sigma}(z) \nonumber \\
    &= pt^2 {\mathcal{G}^*_\sigma (z)}^2 \delta  G_{N-\bar{n}+1\sigma}(z) \, . 
\end{align}
Also, combining this expression with the linearized version of Eq. (\ref{eq_def_Functional}), we finally arrive at the linearized RG equation,
\begin{align}
     \delta G_{N-\bar{n} \sigma}(z) = \lambda_\sigma(z) \delta  G_{N-\bar{n}+1\sigma}(z) \, . 
     \label{eq_linearRG}
\end{align}
where
\begin{align}
     \lambda_\sigma(z) \equiv pt^2 {\mathcal{G}^*_\sigma(z)}^2 \frac{\delta F[\mathcal{G}^*_\sigma(z)]}{\delta \mathcal{G}^*_\sigma } \, .
\end{align}
Here, note that the linearized RG equation (\ref{eq_linearRG}) is written as 
$\delta \tilde{\mu}_{N-\bar{n}}(z) = \lambda_\sigma(z) \delta \tilde{\mu}_{N-\bar{n}+1}(z)$ in terms of the renormalization of the effective chemical potential.

In the deep interior, the fixed point solution for $G^*$ and $\mathcal{G}^*$ has no generation-index dependence.
Then, we have
\begin{align}
     \frac{\delta F[\mathcal{G}_{\sigma}^*(i\omega_l)]}{\delta \mathcal{G}_{\sigma}^* }& = {\mathcal{G}_\sigma^*(i\omega_l)}^{-2} ( \langle {\bar{c}}_{\sigma,l} {c}_{\sigma,l}   \rangle^* )^2 \nonumber \\
     &={\mathcal{G}_{\sigma}^*(i\omega_l)}^{-2}  {G_{\sigma}^*(i\omega_l) }^2 \, , 
\end{align}
where $\langle \cdots \rangle^*$ denotes the expectation value under the fixed point $\mathcal{G}^*$.
In the following, we basically assume a nonmagnetic solution, where $G^*_{\sigma}$ and $\mathcal{G}_{\sigma}^*$ are also independent of $\sigma$.
Thus, the eigenvalue of the RG transformation can be read as
\begin{align}
    \lambda(z) = p[t G_{\sigma}^* (z) ]^2 \, ,
    \label{eq_lambdascaling}
\end{align}
which governs the convergence to the fixed point in the DMFT computation. 
Here, we have omitted $\sigma$ in $\lambda(z)$ for simplicity.
$\lambda(z)$ for the free-electron case is presented in Appendix \ref{appendix_A}.
We will discuss later that $\lambda(z)$ defines the scaling dimension for the RG scheme by setting up an appropriate coordinate associated with a holographic geometry.


\subsection{Operator renormalization}

We next consider a holographic RG-like structure for electron operators attached at the outer edge boundary. 
For the partition function (\ref{eq_def_Z}), we insert an electron operator $c_{N\sigma'\alpha'}(s)$ or 
a composite of $c_{N\sigma'\alpha'}^\dagger(s)c_{N\sigma'\alpha'}(s')$ at the outer edge node of a certain branch in the Bethe lattice network and then investigate how they are renormalized toward the interior.
Of course, the expectation value of a single-electron operator vanishes in the bulk. 
However, such analysis based on the branch Green's function provides an essential view for the holographic relation between the boundary and bulk electrons behind DMFT.

We first consider the RG of a single-electron operator.
Let $\alpha'$ and $\sigma'$ be respectively the branch index and the spin label of the electron attached at an $N$th generation node.
The first step is to resolve
\begin{align}
    &\tilde{g}_{1}(\bar{c}_{N-1},c_{N-1}) \nonumber \\
    &= \int D\bar{c}_{N} Dc_{N} c_{N\sigma'\alpha'}(s) e^{ -\Delta S_{N-1,N} -S_{N}  } \, .
    \label{eq_o_expand}
\end{align}
Similarly to Eq. (\ref{eq_g1perturbation}), we treat $\Delta S_{N-1,N}$ as a perturbation.
Since the expectation value of an odd number of Grassmann variables is zero, the first nontrivial term is
\begin{align}
    &\int d\tau \langle  c_{N\sigma'\alpha'}(s)  \mathcal{H}'(\tau) \rangle_0 \nonumber \\
  &= -t \int d\tau c_{N-1\sigma'\alpha'}(\tau) G_{N\sigma'\alpha'}( s - \tau)
\end{align}
where the term enjoying $(\sigma, \alpha)=(\sigma',\alpha')$ survived.

In general, for the $(2k+1)$th order term in the expansion of Eq. (\ref{eq_o_expand}) with respect to $\Delta S_{N-1,N}\,$, we have ${}_{2k+1}C_{k+1}$ possible selections of $c$ and $\bar{c}$, and then have $(k+1)!$ number of combinations of the brach Green's functions by assuming the free fermionic factorization.
\begin{widetext}
The result can be summarized as 
\begin{align}
& \frac{1}{(2k+1)!} \int d\tau_1\cdots d\tau_{2k+1}  \langle c_{N\sigma'\alpha'}(s)\mathcal{H}'(\tau_1) \cdots \mathcal{H}'(\tau_{2k+1}) \rangle_0 \nonumber \\
& \simeq - \left[ (-)^k\frac{t^{2k} }{k!} 
\sum_{\sigma_1 \cdots \sigma_k} \int d\tau_1 \cdots d\tau_k 
\bar{c}_{N-1\sigma_1}(\tau_1) c_{N-1\sigma_1}(\tau_2) \cdots \bar{c}_{N-1\sigma_k}(\tau_{2k-1}) c_{N-1\sigma_k}(\tau_{2k}) \right. \nonumber \\
& \left. \times
  \sum_{\alpha_1\cdots\alpha_k } G_{\sigma_1\alpha_1}(\tau_1-\tau_2) \cdots G_{\sigma_k\alpha_k}(\tau_{2k-1}-\tau_{2k}) \right] 
\times \left(   t \int d\tau_0 c_{N-1\sigma'}(\tau_0)  G_{N\sigma'\alpha'}(s-\tau_0) \right)
\end{align}
Here, note that the square brackets in the above equation is the same form as Eq. (\ref{eq_2n_expansion_U}). 
In addition,  the branches with $\alpha\ne \alpha'$ give the same contribution as Eq. (\ref{eq_2n_expansion_U}). 
Thus, we have a simple expression
\begin{align}
\tilde{g}_1(\bar{c}_{N-1}, c_{N-1}) &= Z_0 e^{- p t^2 \sum_\sigma \int d\tau_1d\tau_2\bar{c}_{N-1\sigma}(\tau_1)c_{N-1\sigma}(\tau_2) G_{N\sigma}(\tau_1-\tau_2) } 
 \times\Big(t \int d\tau_0 G_{N\sigma'\alpha'}(s-\tau_0)c_{N-1\sigma'}(\tau_0)    \Big) 
\end{align}
which implies the electron operator at the boundary can be renormalized as
\begin{align}
    c_{N\sigma'\alpha'}(s) \to \tilde{c}_{N-1\sigma'}(s) \equiv  t \int d\tau G_{N\sigma'\alpha'}(s-\tau)c_{N-1\sigma'}(\tau)   
    \label{eq_renromalize_c}
\end{align}
After repeating the renormalization process by $\bar{n}$ times,  we obtain
\begin{align}
   & c_{N\sigma'\alpha'}(s) \to   t \int d\tau_{1} G_{N\sigma'\alpha'}(s-\tau_{1}){c}_{N-1\sigma'\alpha_1}(\tau_{1}) \to  t^2 \int d\tau_{1} G_{N\sigma'\alpha'}(s-\tau_{1}) \int d\tau_2 G_{N-1\sigma'\alpha_1}(\tau_1-\tau_2) c_{N-2\sigma'\alpha_2}(\tau_{2}) \nonumber \\
    &\to \cdots     
    \to  
    t^n\int d\tau_{1}d\tau_2 \cdots d\tau_{\bar{n}} G_{N\sigma'\alpha'}(s-\tau_{1})G_{N-1\sigma'\alpha_1}(\tau_1-\tau_2) \cdots G_{N-\bar{n}+1\sigma'\alpha_{\bar{n}-1} }(\tau_{\bar{n}-1}- \tau_{\bar{n}})c_{N-\bar{n}\sigma'\alpha_{\bar{n}}}(\tau_{\bar{n}})  \, ,
\label{eq_oprg_c}
\end{align}
where we have recovered the branch indices $\{ \alpha_1, \alpha_2 \cdots \alpha_{\bar{n}} \}$ specifying the branch where the edge electron operator is attached.  

Eq. (\ref{eq_oprg_c}) is of the convolution form, so that we have a simple expression of the operator renormalization in the Fourier mode representation,
\begin{align}
         t\int d\tau &G_{N\sigma'\alpha'}(s-\tau)c_{N-1\sigma'}(\tau) 
         = \frac{1}{\sqrt{\beta}}\sum_l e^{-i\omega_l s}  c_{N-1\sigma',l} t G_{N\sigma'\alpha'}(i\omega_l)\, ,
\end{align}
implying
\begin{align}
    &c_{N\sigma'\alpha',l} \to  t G_{N\sigma'\alpha'}(i\omega_l) c_{N-1\sigma'\alpha_1,l} \to \cdots 
    \to  t G_{N\sigma'\alpha'}(i\omega_l) \cdots   tG_{N-\bar{n}+1\sigma'\alpha_{\bar{n}-1}}(i\omega_l)c_{N-\bar{n}\sigma'\alpha_{\bar{n}},l}\, .
    \label{eq_frg_fourier}
\end{align}
In the same way, we also have 
\begin{align}
    &\bar{c}_{N\sigma'\alpha_1,l} \to \cdots 
    \to  tG_{N\sigma'\alpha_1}(i\omega_l) \cdots  tG_{N-\bar{n}\sigma'\alpha_{\bar{n}}}(i\omega_l)\bar{c}_{N-\bar{n}\sigma'\alpha_{\bar{n}},l}\, .
    \label{eq_fcrg_fourier}
\end{align}

When assuming the fixed-point branch Green's function at the outer edge nodes, i.e. $G_N=G^*$, the branch Green's function becomes uniform everywhere on the Bethe lattice network, independently of the branch index $\{\alpha\}$.
Then, we can express the renormalized electron operators as 
\begin{align}
    {c}_{N\sigma',l}     &\to [tG^*_{\sigma'}(i\omega_l)]^{\bar{n}}     {c}_{N-\bar{n}\sigma',l} \nonumber\\
    \bar{c}_{N\sigma',l} &\to [tG^*_{\sigma'}(i\omega_l)]^{\bar{n}} \bar{c}_{N-\bar{n}\sigma',l}
    \label{eq_operator_RG}
\end{align}
where we have dropped the $\alpha$ index again.

We next consider the renormalization of the composite operator $c_{N\sigma'\alpha'}^\dagger(s)c_{N\sigma'\alpha'}(s')$, which is more relevant to a holographic argument in the next section.
In the following, we assume $s>s'$ for simplicity.
Note that $c_{N\sigma'\alpha'}^\dagger(s)c_{N\sigma'\alpha'}(s-\epsilon)$ with $\epsilon=+0$ reduces to the electron number operator at the outer edge node.
After similar calculations above, we obtain 
\begin{align}
    & \bar{c}_{N\sigma'\alpha'}(s)c_{N\sigma'\alpha'}(s') \nonumber \\
    &\to  G_{N\sigma'\alpha'}(s'-s) + t^2\int d\tau_1 d\tau_2 \bar{c}_{N-1\sigma'\alpha_1}(\tau_1)c_{N-1\sigma'\alpha_1}(\tau_2) G_{N\sigma'\alpha'}(\tau_1-s)G_{N\sigma'\alpha'}(s'-\tau_2) \, ,
    \label{eq_n_1ststep}
\end{align}
where $\alpha_1$ is the branch index for the $(N-1)$th generation.
Here, note that the expectation value of $\langle \bar{c}_{N\sigma'\alpha'}(s)c_{N\sigma'\alpha'}(s')\rangle$ with respect to the entire
Bethe lattice system is different from $ G_{N\sigma'\alpha'}(s'-s)$, which is the branch Green function defined for the descendant branch disconnected from the parent node.
The first term in Eq. (\ref{eq_n_1ststep}) represents the contribution from the parent node to the bulk expectation value.    
Using Eq. (\ref{eq_n_1ststep}) recursively, we can further renormalize the electron density operator as
\begin{align}    
    &\to  G_{N\sigma'\alpha'}(s'-s) + t^2\int d\tau_1 d\tau_2 G_{N-1\sigma'\alpha_1}(\tau_1-\tau_2)  G_{N\sigma'\alpha'}(\tau_1-s)G_{N\sigma'\alpha'}(s'-\tau_2) 
   \nonumber \\
    &+  t^4 \int d\tau_1 \cdots d\tau_4 \bar{c}_{N-2\sigma'\alpha_2}(\tau_3)c_{N-2\sigma'\alpha_2}(\tau_4) G_{N-1\sigma'\alpha_1}(\tau_3-\tau_1) G_{N\sigma'\alpha'}(\tau_1-s)G_{N-1\sigma'\alpha_1}(\tau_2-\tau_4)G_{N\sigma'\alpha'}(s'-\tau_2)   \nonumber  \\
    & \to \cdots \, ,\label{eq_n_2ndstep}
\end{align}
where $\alpha_n$ is the branch index for upper branches connected to $\alpha'$.

Turning to the Fourier space, we also have the expression after $\bar{n}$-times renormalization as
\begin{align}
& \bar{c}_{N\sigma'\alpha',l} c_{N\sigma'\alpha',l'}\nonumber \\
& \to     [ G_{N\sigma'\alpha'}(\omega_l)+\cdots+ t^{2(\bar{n}-1)} G^2_{N\sigma'\alpha'}(\omega_l)\cdots G^2_{N-\bar{n}+2\sigma'\alpha_{\bar{n}-2}}(i\omega_l)  G_{N-\bar{n}+1\sigma'\alpha_{\bar{n}-1}}(i\omega_l)] \delta_{l,l'}
   \nonumber \\
   &+   
   t^{2\bar{n}} [G_{N\sigma' \alpha'}(i\omega_l)\cdots G_{N-\bar{n}+1\sigma'\alpha_{\bar{n}-1}}(i\omega_l)\bar{c}_{N-\bar{n}\sigma'\alpha_{\bar{n}},l}][ G_{N\sigma' \alpha'}(i\omega_{l'})  \cdots G_{N-\bar{n}+1\sigma'\alpha_{\bar{n}-1}}(i\omega_{l'})  c_{N-\bar{n}\sigma'\alpha_{\bar{n}},l'}]\, .
\end{align}
Note that the 2nd line terms represent to the contribution to the bulk expectation value for $\bar{n} \gg 1$, while the 3rd line terms corresponds to the renromalized electrons of Eqs. (\ref{eq_frg_fourier}) and (\ref{eq_fcrg_fourier}). 
Assuming the fixed-point branch Green's function independent of $\alpha$ at the outer edge nodes, we arrive at 
\begin{align}
 & \bar{c}_{N\sigma'\alpha',l} c_{N\sigma'\alpha',l'} \nonumber \\ 
  &\to    G^*_{\sigma'}(i\omega_l)\frac{ 1- [t G^*_{\sigma'}(i\omega_l)]^{2\bar{n}}}{1 - [t G^*_{\sigma'}(i\omega_l)]^2}\delta_{l,l'}
   +  
   [t G^*_{\sigma'}(i\omega_l)]^{\bar{n}}\bar{c}_{N-\bar{n}\sigma',l} [t G^*_{\sigma'}(i\omega_{l'}) ]^{\bar{n}}  c_{N-\bar{n}\sigma',l'} \, . 
\end{align}
Here, we should note that 
\begin{align}
    G^\mathrm{o}_{\sigma'}(i\omega_l)  \equiv \frac{ G^*_{\sigma'}(i\omega_l) }{1 - [t G^*_{\sigma'}(i\omega_l)]^2}
    \label{dq_bulk_density}
\end{align}
corresponds to the bulk Green's function.
Subtracting this as a background, we can thus read the renormalization of the electron composite operator as
\begin{align}
 & \left(\bar{c}_{N\sigma'\alpha',l} c_{N\sigma'\alpha',l'} -   G^\mathrm{o}_{\sigma'}(i\omega_l) \delta_{l,l'}\right)  
   \to  
   [t G^*_{\sigma'}(i\omega_l)]^{\bar{n}}   [t G^*_{\sigma'}(i\omega_{l'})]^{\bar{n}}  \left( \bar{c}_{N-\bar{n}\sigma',l} c_{N-\bar{n}\sigma',l'}  -  G^\mathrm{o}_{\sigma'}(i\omega_l)\delta_{l,l'} \right)\, .
   \label{eq_density_cor_rg}
\end{align}

\end{widetext}

\subsection{Boundary correlation functions}

Using the operator renormalization, we can calculate the correlation functions of a pair of operators at the outer edge boundary under the bulk-fixed-point condition.
We write the outer edge generation index as $N$ and the branch indices for the operator pair as $\alpha$ and $\alpha'$ ($\alpha \ne \alpha'$). 
Here, it is important to note that the operator renormalization in the previous subsection can be performed independently for the two branches of operators before they are renormalized to fuse with each other.
Suppose that the two operators meet after $\bar{n}(\alpha,\alpha')$ times of renormalization, i.e., at the generation index $N-\bar{n}(\alpha,\alpha')$.
We then obtain the boundary electron correlation function as 
\begin{align}
& G_{N\alpha\alpha}(\tau-\tau') = - \langle c_{N\sigma\alpha} (\tau) \bar{c}_{N\sigma\alpha'}(\tau') \rangle
 \nonumber \\
 &=-\frac{1}{\beta}\sum_{ll'} e^{-i\omega_l \tau + i \omega_{l'}\tau'} \langle c_{N\sigma\alpha,l}\bar{c}_{N\sigma\alpha',l'} \rangle  \nonumber \\
 &=-\frac{1}{\beta}\sum_{l} e^{-i\omega_l (\tau -\tau')}\left[t G^*(i\omega_l)\right]^{2\bar{n}}\langle c_{N-\bar{n}\sigma,l}\bar{c}_{N-\bar{n}\sigma,l}\rangle \, ,
\end{align}
for $0\le \tau - \tau' <\beta$.
\footnote{$\langle \cdots \rangle$ denotes the bulk expectation value for the entire Bethe lattice network, which is distinct from $\langle \cdots \rangle^*$ with the fixed point action including the descendant branch only.}
Note that the bulk expectation value is independent of $\bar{n}$ and $\alpha$, as well as $\sigma$, if there is no magnetic order. 
Moreover, the bulk correlation function is given by Eq. (\ref{dq_bulk_density}): 
\begin{align}
         \langle c_{N-\bar{n}\sigma,l}\bar{c}_{N-\bar{n}\sigma,l}\rangle = - G^\mathrm{o}(i\omega_l) 
         \, ,
\end{align}
which reduces to the branch Green's function for $p \gg 1$.
Thus, we obtain 
\begin{align}
     & G_{N\alpha\alpha'}(\tau-\tau')  
     \nonumber \\
      &= \frac{1}{\beta t} \sum_{l} e^{-i\omega_l (\tau -\tau')} \left[t G^*(i\omega_l)\right]^{2\bar{n}} G^\mathrm{o}(i\omega_l)  \nonumber \\
    & \sim \sum_l e^{-i\omega_l (\tau-\tau')} G^\mathrm{o}(i\omega_l) e^{ - 2\bar{n}\log (1/ [t G^*(i\omega_l)]) } \, 
    \label{eq_eqt_cor}
\end{align}
for $\bar{n}\gg1$, which implies that the asymptotic behavior is governed by the maximum of $|G^*(i\omega_l)|$.
Here, note that the $\bar{n}$ dependence in Eq. (\ref{eq_eqt_cor}) is decoupled from the $\tau$-dependent term since $G^*(i\omega_l)$ is always a local Green's function in the DMFT framework.

For example, the branch Green's function for the free electron with $\mu=0$ is explicitly calculated as  
\begin{align}
G^*(i\omega_l) = i\frac{\omega_l \mp \sqrt{\omega_l^2 +4pt^2}}{2pt^2} \, ,   
\end{align}
where the $\mp$ sign corresponds to positive and negative $\omega_l$, respectively (see Appendix \ref{appendix_A}).
Then, the maximum of $G^*(i\omega_l)$ is specified by $\omega^*\equiv \omega_{l^*}$ with $l^*=0, -1$, and thus the asymptotic form of the boundary correlation function for $\bar{n}(\alpha,\alpha') \gg 1$ behaves as follows.
\begin{align}
    &G_{N\alpha\alpha'}(\tau-\tau') 
    \sim \sin (\frac{\pi (\tau -\tau')}{\beta} ) \exp( - \frac{2 \bar{n}}{\xi} ) \, ,
     \label{eq_corr_asympt}
\end{align}
where
$ 1/\xi = \log \left(\frac{\sqrt{ \pi^2+4p(\beta t)^2 } - \pi}{2 p \beta t}\right)$.

In a similar way, we can also consider the correlation function of the composite operator, which is relevant to a holographic argument in the context of the response function against a perturbation with respect to $\tilde{\mu}(\omega)$ in the next section.
Taking into account Eq. (\ref{eq_density_cor_rg}), we analyze the correlation function of the normal ordered operators
\begin{align}
D_{N\alpha\alpha'}(\tau - \tau')\equiv 
\langle &[\bar{c}_{ N\sigma\alpha^{\;}}(\tau\;)c_{N\sigma\alpha^{\;}} (\tau\;)   - G_{\sigma}^\mathrm{o}(0\;)]\nonumber \\
        &[ \bar{c}_{N\sigma\alpha'}(\tau')c_{N\sigma\alpha'} (\tau') -  G^\mathrm{o}_{\sigma}(0) ]\rangle 
 \end{align}
for $0\le \tau - \tau' <\beta$, where we have dropped the $\sigma$ index in $D_{N\alpha\alpha'}$ assuming a non-magnetic phase. 

 Let the two boundary nodes meet after $\bar{n}(\alpha,\alpha')$ times renormalization.
 We then have
\begin{align}
& D_{N\alpha\alpha'}(\tau - \tau') \nonumber \\
 &=\frac{1}{\beta^2}\sum_{\{l\}} e^{i(\omega_{l_1} - \omega_{l_2} )\tau +i(\omega_{l_3} - \omega_{l_4} )\tau'}  \nonumber \\
 & \hspace{1cm} \times \langle [\bar{c}_{N\sigma\alpha,l_1}c_{N\sigma\alpha,l_2}-G^\mathrm{o}(i\omega_{l_1})\delta_{l_1,l_2} ]\nonumber \\
 & \hspace{1.5cm}   [\bar{c}_{N\sigma\alpha',l_3}{c}_{N\sigma\alpha',l_4} - G^\mathrm{o}(i\omega_{l_3})\delta_{l_3,l_4}] \rangle  \nonumber \\
 &=\frac{1}{\beta^2}\sum_{\{l\}}  e^{i(\omega_{l_1} - \omega_{l_2} )\tau +i(\omega_{l_3} - \omega_{l_4} )\tau'}  \Gamma^\mathrm{o}_{l_1,l_2,l_3,l_4} \nonumber \\
& \times  \left[t G^*(i\omega_{l_1}) t G^*(i\omega_{l_2})t G^*(i\omega_{l_3})t G^*(i\omega_{l_4})\right]^{\bar{n}} \, ,
\label{eq_D_cor}
\end{align}
where
\begin{align}
\Gamma^\mathrm{o}_{l_1,l_2,l_3,l_4} = \langle& [\bar{c}_{N-\bar{n}\sigma,l_1}c_{N-\bar{n}\sigma,l_2}-G^\mathrm{o}(i\omega_{l_1})\delta_{l_1,l_2} ]\nonumber \\
 &  [\bar{c}_{N-\bar{n}\sigma,l_3}{c}_{N-\bar{n}\sigma,l_4} - G^\mathrm{o}(i\omega_{l_3})\delta_{l_3,l_4}] \rangle \,.
\end{align} 
Since $\Gamma^\mathrm{o}_{l_1,l_2,l_3,l_4}$ is independent of the network distance $\bar{n}(\alpha,\alpha')$, the long-distance behavior of $D_{N\alpha,\alpha'}$ is also governed by the maximum of $|tG^*(i\omega_{l})|$ as in the case of Eq. (\ref{eq_eqt_cor}).
Suppose that $\omega^*$ corresponds to $l_1=l_2 =l_3 = l_4 = l^*$.
We then have the asymptotic behavior of the amplitude 
\begin{align}
 D_{N\alpha\alpha'}  & \sim \Gamma^\mathrm{o}_{l^*,l^*,l^*,l^*} 
\left[ t G^*(i\omega^*) \right]^{4 \bar{n}} \nonumber \\
 & \sim e^{-4 \bar{n} \log(1/[t G^*(i\omega^*)])}\, ,
\label{eq_D_asympto}
\end{align}
which is also decoupled from the time-dependent term.

\section{Scaling dimensions in a holographic network}
\label{sec_3}

So far we have discussed the recursive RG transformation on the lattice.
In order to discuss the relation between Eq. (\ref{eq_lambdascaling}) and Eq. (\ref{eq_D_asympto}) in the context of holography, we need to clarify the geometry inherent in the Bethe lattice network.

As discussed for the Bethe lattice Ising model~\cite{Oku2023}, we can set up a disk-like coordinate, where the radial coordinate with a generation index $n$ is defined as
\begin{align}
    R_n=1-r_n
    \label{eq_radiusR}
\end{align}
with
\begin{align}
    r_n\equiv p^{-n}\, .
    \label{eq_radiusr}
\end{align}
Here, $r_n$ measures the distance toward the interior from the outer edge boundary of the unit disk ($r_n \to 0$ for $n, N\to\infty$).
Note that the circles in the base plane of the cylinder in Fig.~\ref{tag_fig1} are illustrated by Eq. (\ref{eq_radiusR}).
On the other hand, the distance between the two nodes on the outer edge boundary along the circumference direction is defined as
\begin{align}
x(\alpha,\alpha')=p^{\bar{n}(\alpha,\alpha')}\, .
\label{eq_cdistance}
\end{align}
where $\alpha$ and $\alpha'$ denote branch indices for the two outer edge nodes (see Fig.~\ref{tag_fig1}).
Here, note that Eq.~(\ref{eq_cdistance}) is consistent with the $p$-adic norm\cite{Brekke1993} in the $p$-adic AdS/CFT.

As in Ref. [\onlinecite{Oku2023}], we further introduce an effective coordinate by smearing out the network nodes so that each lattice point has a unit area.
As $n$ changes with $n\to n+1$, the length increases by one. 
Also, the length of the circle at a fixed value of the radius $R_n=1-r_n$ is $qp^{n-1}$. 
Then the effective coordinate can be read as
\begin{align}
ds^2 = \frac{L^2}{r^2}(dr^2+ d x^2 )\, .
\label{eq_ads2}
\end{align}
as $r\to 0$, where $x$ represents the rescaled coordinate,  $ \frac{q\log p}{2\pi p} x \to x $, along the outer edge circle of Fig. \ref{tag_fig1}.
For $n \gg 1$, this metric describes the Poincar\'e upper half with the radius $L\equiv 1/\log p$ in the non-compact limit.
Here, it is essential to note that Eq. (\ref{eq_ads2}) does not explicitly contain $d\tau$, since it was introduced after completing the imaginary-time path integral.

Then, the asymptotic behavior of the boundary correlation function can be expressed as
\begin{align}
     G_{N\alpha\alpha'} 
     &\sim e^{ - 2(N-n)\log |1/t G^*(i\omega^*)| } \sim x^{-2\Delta_G } \,.
     \label{eq_scale_corr}
\end{align}
with the scaling dimension,
\begin{align}
    \Delta_G \equiv -\frac{\log |tG^*(i\omega^*)|}{\log p}
    \label{eq_scaling_dim}
\end{align}
where $\omega^*$ is defined as the frequency $\omega_l$ where $|G^*|$ takes its maximum value.
Also, from Eq. (\ref{eq_operator_RG}), it follows that 
\begin{align}
    {c}_{n\sigma'}(i\omega^*) &= [tG^*(i\omega^*)]^{n-N}     {c}_{N\sigma'}(i\omega^*) \nonumber\\
    &\sim r_n^{\Delta_G}  {c}_{N\sigma'}(i\omega^*) \, ,
\end{align}
which is consistent with Eq. (\ref{eq_scale_corr}).
Since $tG^*(i\omega^*)$ is a pure imaginary with $|tG^*(i\omega^*)|<1$, $\omega^*$ naturally corresponds to the slowest decay mode.
We explicitly have $\omega^* = \omega_{l=0}$ for the free electron case.

In addition, the Fourier decomposition of Eq. (\ref{eq_eqt_cor}) defines the descendant spectrum of the scaling dimension
\begin{align}
\Delta_{G}^{(l)} =  -\frac{\log |tG^*(i\omega_l)|}{\log p}
\label{eq_desc_scaling_dim}
\end{align}
for $\omega_l$ around $\omega^*$. 
Here, we should recall that the Bethe lattice Ising model, which is completely a classical model, has no descendant spectrum.
Thus, $\Delta_G^{(l)}$ demonstrates the quantum nature in DMFT, originating from the imaginary time path integral.
In the metallic phase, the distribution of the quantum number $l$ is the same as in the free-electron case.
However, the nontrivial interaction effect is built into the fixed-point Green's function $G^*(i\omega_l)$.

We further illustrate the asymptotic behavior of the correlation function $D_{N\alpha\alpha'}$ in the effective coordinate.
From Eq. (\ref{eq_D_asympto}), it straightforwardly follows that 
\begin{align}
   D_{N\alpha\alpha'}   &\sim e^{ - 4\bar{n}\log [1/t G^*(i\omega^*)] } \sim x^{-2\Delta_D } \,.
     \label{eq_scale_corr_D}
\end{align}
with the scaling dimension,
\begin{align}
    \Delta_D \equiv 2 \Delta_G= -\frac{2\log |tG^*(i\omega^*)|}{\log p}
    \label{eq_scaling_dim_D}
\end{align}
This simple relation reflects the fact that the decay rate of the correlation function is determined by the number of renormalized electron operators included in the composite operator.
In addition to the slowest mode, we can extract the descendant spectrum from Eq. (\ref{eq_D_cor}), 
\begin{align}
\Delta_{D}^{(l_1l_2l_3l_4)} = -\frac{\sum_{\nu=1}^4 \log |tG^*(i\omega_{l_\nu})|}{\log p}  
\label{eq_desc_scaling_Ddim}
\end{align}
Here, we note that the precise spectrum should be specified by non-zero elements of vertex function
$
\Gamma^\mathrm{o}_{l_1,l_2,l_3,l_4}
$
at the bulk node.

The above scaling dimension for $D_{N\alpha\alpha'}$ has an interesting connection to the linearized RG for Green's function.
As discussed in subsection \ref{subsec_3-C}, a perturbation for the bulk solution,   $G^*+\delta G$,  at an outer edge node penetrates the interior with $\delta G_{n}(z) = \lambda(z)^{N-n} \delta G_{N}(z)$.
Using the radial coordinate $r_n$, we obtain
\begin{align}
     \delta G_{n}(z) \propto \lambda(z)^{N-n} \propto r_n^{1-\Delta_D}\, ,
\label{eq_scale_G}
\end{align}
which implies that the branching structure of the Bethe lattice network is balanced with the scaling property of the $r$ coordinate (\ref{eq_radiusr}).
Note that $p^N$ involved in Eqs. (\ref{eq_scale_G}) corresponds to the cutoff scale.
For the free-electron case, we explicitly have
\begin{align}
    \Delta_D = -\frac{2\log \left( \frac{\sqrt{\pi^2+4p(\beta t)^2} -\pi}{2p\beta t}  \right)}{\log p} \to 1
\end{align}
for $\beta\to \infty$ or $p\to \infty$.

Here, we discuss the relationship of Eqs. (\ref{eq_scale_G}) and (\ref{eq_scale_corr_D}), referring to a massive scalar field $\phi$ in AdS$_{d+1}$, where $d$ is the dimension of a boundary CFT.
In AdS$_{d+1}$, the scaling property near the AdS boundary ($r \to 0$) is described by
\begin{equation}
\phi(r) \sim A r^{\Delta_-} + B r^{\Delta_+} \, , 
\label{eq_adscft}
\end{equation}
where $A$ represents a boundary field coupled to the boundary CFT, and $B$ gives a response of the operator against a perturbation with respect to $A$.
Then, it is established that the scaling dimensions in Eq. (\ref{eq_adscft}) enjoy $\Delta_+ + \Delta_- =d$.
From Eqs. (\ref{eq_scale_corr_D}) and (\ref{eq_scale_G}), the scaling dimensions in DMFT corresponding to $A$ and $B$ fields can be interpreted as $\Delta_+ \equiv 1- \Delta_D$ and $\Delta_- \equiv \Delta_D$, respectively.
Moreover, these scaling dimensions clearly satisfy the scalar-field relation,
\begin{align}
    \Delta_+ + \Delta_- =1 \, ,
    \label{eq_adsdim}
\end{align}
with $d=1$.
Here, note that $d=1$ is characteristic of the tree network nature of the Bethe lattice, or, equivalently, the Burhat-Tits tree, which is consistent with the Bethe lattice Ising model~\cite{Oku2023}, as well as the $p$-adic AdS/CFT~\cite{Gubser2017}.
The relation (\ref{eq_adsdim}) would reflect that the branch Green's function for DMFT is a ``classicalized' quantity in the sense of the expectation value on the loop-free network.
For a thorough understanding of the quantum effect in Eq. (\ref{eq_adsdim}), it would be essential to verify the many-body correlation effect at the level of the Bethe lattice Hubbard model beyond the effective medium in DMFT.\cite{Lepetit2000,Eckstein2005,Lunts2021,Chen2025}
Meanwhile, the excitation spectra of $\Delta_G^{(l)}$ and $\Delta_D^{(l_1l_2l_3l_4)}$ clearly illustrate the quantum effect through the imaginary-time path integral.

In addition, we should comment on the role of the spatial dimension in DMFT, which corresponds to the coordination number $q(=p+1)$ in the Bethe lattice network.
A crucial point is that the branching number $p$ is associated with Poincar\'e radius $L=1/\log p$ rather than $d=1$ corresponding to the dimension of the boundary theory in AdS$_{d+1}$/CFT$_{d}$.
Thus, DMFT corresponds to the large curvature limit in the context of AdS$_2$, although it is justified in the infinite spatial dimension.
In the context of $p$-adic AdS/CFT~\cite{Gubser2017}, several possible field extensions were discussed to implement ``spatial dimension" $d>1$ by modifying the definition of the norm for the boundary field theory side. 
However, their relevance to DMFT is not yet clear, which may be an interesting future issue.

\section{Numerical results}
\label{sec_numerical}

To verify the scaling dimensions for interacting electrons, we apply DMFT to the Hubbard model on the Bethe lattice [Eq. (\ref{H})]. 
Following the convention of DMFT, let us introduce the normalized coupling as $\tilde{t}= t\sqrt{p} $ and then define the half bandwidth, $\tilde{D}\equiv 2 \tilde{t}$.
In the noninteracting case, the density of states has a semicircular form,
$\rho(\omega)=2/(\pi \tilde{D})\sqrt{1-(\omega/\tilde{D})^2}$.
In the following, we set $\tilde{D}$ as an energy unit.
In DMFT, the self-consistent equation is given as,
\begin{align}
{\cal G}^{-1}(i\omega_l)=i\omega_l+\mu-\left(\frac{\tilde{D}}{2}\right)^2G(i\omega_l),\label{self}
\end{align}
where ${\cal G}$ is the non-interacting Green's function for the effective impurity model.
Of course, this self-consistent equation is essentially equivalent to the fixed-point equation of Eq.~(\ref{FPRR}).
Given non-interacting Green's function ${\cal G}$, we numerically obtain Green's function ${G}$ by means of the impurity solver, update ${\cal G}$ [Eq.~(\ref{self})], and iterate this self-consistent loop until the result converges within numerical accuracy. 
Then, the fixed-point solution of $G^*(\tau)$ can be obtained in the framework of DMFT.

For a non-magnetic and half-filled system $(\mu=U/2)$, it is well established that strong Coulomb interactions lead to a Mott transition. 
To quantitatively discuss how the scaling dimensions vary with increasing interaction strength, we employ the hybridization-expansion continuous-time quantum Monte Carlo (CTQMC) method~\cite{Werner_2006,LoopQMC_imp2011} based on the segment algorithm~\cite{Werner_2006,Koga_2011} as the impurity problem solver.
In our simulations, we fix the inverse temperature as $\beta=100$ and set the number of data points along the imaginary time direction to $N_\tau =1000$.

\begin{figure}
    \centering
    \includegraphics[width=0.7\linewidth]{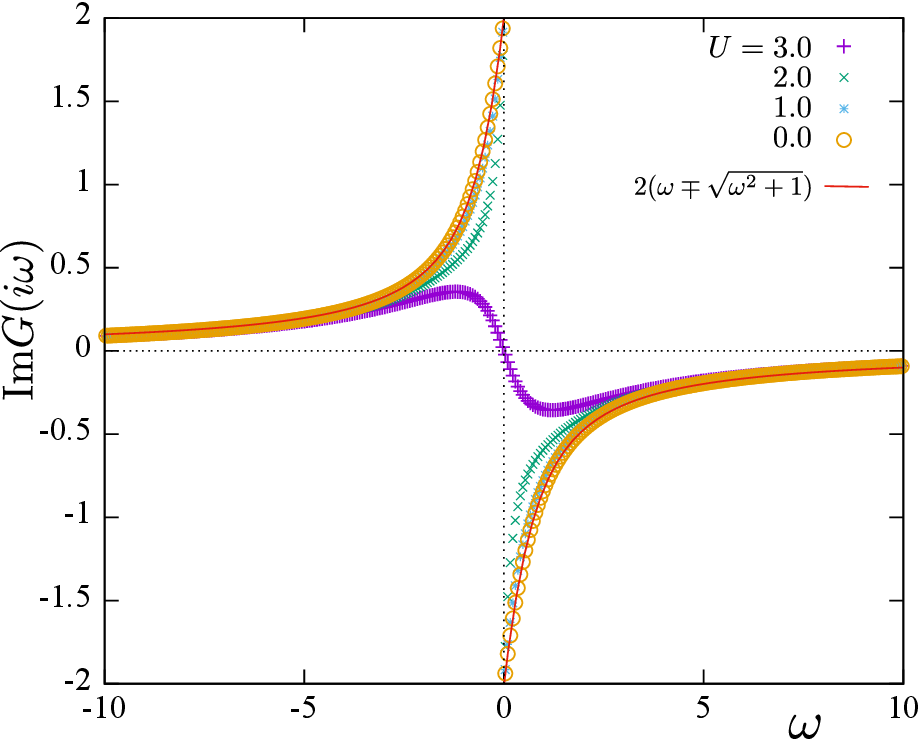}
    \caption{Imaginary part of $G^*(i\omega)$ for $U=0.0$, 1.0, 2.0, and 3.0. 
    Solid curve represents the free electron curve $\mathrm{Im} G^*(i\omega) = 2(\omega \mp \sqrt{\omega^2 +1})$.}
    \label{tag_fig3}
\end{figure}

 For $\omega_l = \frac{\pi(2l+1)}{\beta}$ for $l=0, \pm 1, \cdots$, then we discretize $G^*(\tau)$ into 
\begin{align}
       G^*(i\omega_l)  &\simeq \frac{\beta}{N_\tau}\sum_{n=0}^{N_\tau-1}
G^*\left(\frac {\beta}{N_\tau} n\right) e^{i\frac{\pi(2l+1)}{N_\tau} n} \, , \label{eq_discreteG}
\end{align}
the imaginary part of which is presented in Fig.~\ref{tag_fig3}.
In the figure,  $|\mathrm{Im} G^*| \to 2$ as $\omega\to  0$ for $U=0, 1.0$ and $2.0$, indicating the metallic phase.
 Indeed, the curve for $U=0$ is consistent with the analytic expression of Green's function with $ D= 2\tilde{t}=1$ for the free electron;
\begin{align}
    G^*(i\omega) =  \frac{2i}{\tilde{D}}(\omega\mp\sqrt{\omega^2 + \tilde{D}^2} ) =2i(\omega \mp \sqrt{\omega^2 +1})\, ,
\end{align}
which is illustrated as a solid line in Fig.~\ref{tag_fig3}.
For $U=3.0$ and $\beta=100$, on the other hand, $|\mathrm{Im} G^*| $ exhibits a maximum at $\omega^* \simeq \pm 1.23 $.
The Mott transition in DMFT is of the first order.
Then, the transition points for $\tilde{D}=1.0$ at $\beta=100$ are estimated as $U_{c1}\simeq 2.36$ and $U_{c2}\simeq 2.55$, between which the metallic and insulating solutions may coexist. 
We also note that the transition point at $T=0$ is also estimated as $U_{c2} \simeq 2.94 \tilde{D}$ in Ref. [\onlinecite{Bulla1999}].
Fig.~\ref{tag_fig3} is consistent with these known results.

\begin{figure}
    \centering
    \includegraphics[width=0.7\linewidth]{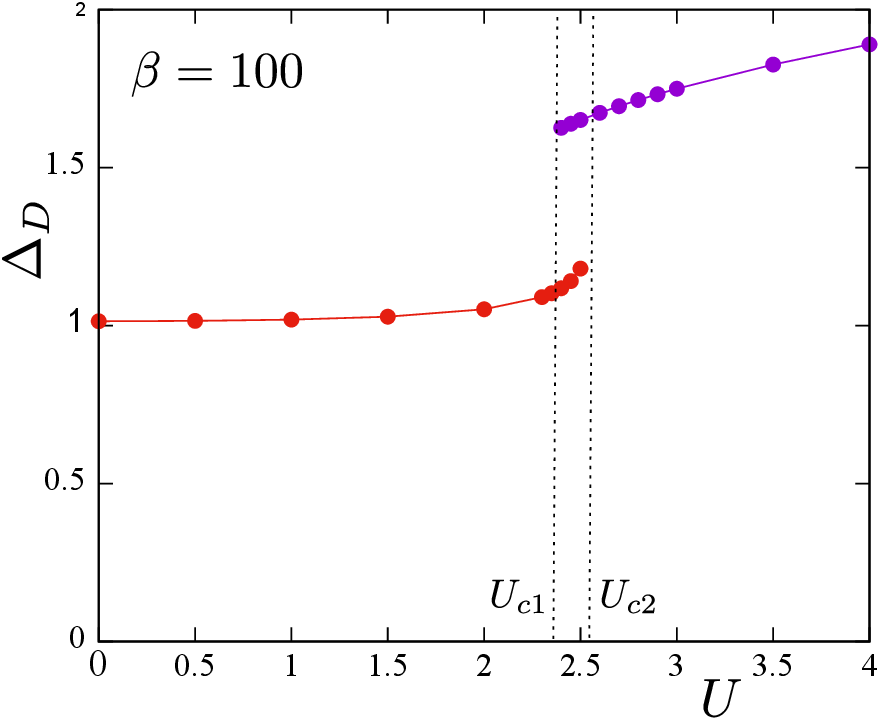}
    \caption{The scaling dimension $\Delta_D$ for $p=100$ at $\beta=100$.
    The lower branch indicates the metallic solution, while the upper branch corresponds to the insulating one.
    The vertical dotted lines represent $U_{c1}\simeq 2.36$ and $U_{c2}\simeq 2.55$.}
    \label{tag_fig4}
\end{figure}

Figure \ref{tag_fig4} shows the scaling dimension $\Delta_D$ for $\tilde{D}=1$ with $p=100$ at $\beta=100$, which is a sufficiently low temperature.
In order to analyze the scaling dimension in association with DMFT, it is useful to rewrite the scaling dimension (\ref{eq_scaling_dim}) into 
\begin{align}
\Delta_D  
 = -\frac{2\log( \tilde{t} G^*(i\omega^*) /\sqrt{p} )}{\log p}  
& = 1 - \frac{2\log \tilde{t} G^*(i\omega^*)}{\log p}\, .
\label{eq_Delta}
\end{align}
For the metallic case ($U < U_c$), we can then observe 
\begin{align}
    \Delta_D \simeq 1\, . 
    \label{eq_delta_ff}
\end{align}
This result implies that the metallic phase realized in DMFT corresponds to a Fermi liquid fixed point, which is characterized by $|G^*(i\omega)|=2$ as $\omega \to 0$.
Indeed, we can confirm $\Delta_D  \simeq 1 $ up to $U\simeq 2.5$ in Fig. \ref{tag_fig4}.
For the insulating phase ($U>U_c$), on the other hand, $\omega^*$ deviates from $\omega=0$, at which $G^*(i \omega^*)<2$. 
In Fig.~\ref{tag_fig4}, $\Delta_D$ actually has a gap between the free-fermion value $\Delta_D\simeq 1$ and about $\Delta_D\sim 1.7 $ for $U_{c1} <U <U_{c2}$.
This gap is consistent with the fact that the Mott transition is of the first order within the DMFT calculation\cite{Werner2007}.

Here, it should be noted that the free scalar field (\ref{eq_adscft}) in AdS/CFT corresponds to the $\omega=0$ mode.
In contrast, the lowest-energy mode of the Bethe lattice Hubbard model in the insulating phase clearly shifts from $\omega=0$ to $\omega^*\simeq 1.23$, implying that $\Delta_D$ and $\Delta_G$ reflect an essential difference of the interior induced by the interaction $U$ between the metallic and insulating phases.  
On the other hand, the relation (\ref{eq_adsdim}) holds for both phases, suggesting that this relation originates from the network structure independent of $U$.

Next, we analyze the excitation spectrum of $\Delta_D^{(l)}$.
As mentioned below Eq. (\ref{eq_desc_scaling_Ddim}), we need nonzero elements of
\begin{align}
\Gamma^\mathrm{o}_{l_1,l_2,l_3,l_4} = &\langle \bar{c}_{N-\bar{n}\sigma,l_1}c_{N\sigma,l_2}
\bar{c}_{N-\bar{n}\sigma,l_3}{c}_{N-\bar{n}\sigma,l_4} \rangle \nonumber \\
&-G^\mathrm{o}(i\omega_{l_1})G^\mathrm{o}(i\omega_{l_3})\delta_{l_1,l_2}\delta_{l_3,l_4}\,.
\end{align}
However, it is a hard task to compute the 4-body term for the effective impurity problem.
Here, assuming the free fermionic factorization, we obtain $\Gamma^\mathrm{o}_{l_1,l_2,l_3,l_4}$ at the level of the effective medium in DMFT:  
\begin{align}
\Gamma^\mathrm{o}_{l_1,l_2,l_3,l_4} \simeq - G^\mathrm{o}(i\omega_{l_1})G^\mathrm{o}(i\omega_{l_3})\delta_{l_1,l_4}\delta_{l_2,l_3}\, ,
\end{align}
yielding
\begin{align}
\Delta_{D}^{(l,l')} &= -2\frac{\log |tG^*(i\omega_{l})|+\log |tG^*(i\omega_{l'})|}{\log p}  \nonumber \\ 
&= 2\left(1- \frac{\log |\tilde{t}G^*(i\omega_{l})|+\log |\tilde{t}G^*(i\omega_{l'})|}{\log p} \right) 
\label{eq_desc_scaling_Ddim2}
\end{align}
for $l, l'$ around $l^*$.

Figure 5 shows the spectra for $U=1.0$(metallic phase) and $U=3.0$ (insulating phase),
 where $\Delta_{D}^{(l,l')} $ is plotted against
\begin{align}
    l_s \equiv l+l' -2 l^* \, .
\end{align}
In the figure, the vertical axis is adjusted so that $\Delta_D$ corresponds to the origin.
Note that $\Delta_{D}^{(l,l')} $ illustrates the contributions only from $l, l' \ge 0$.

\begin{figure}
    \centering
    \includegraphics[width=1.0\linewidth]{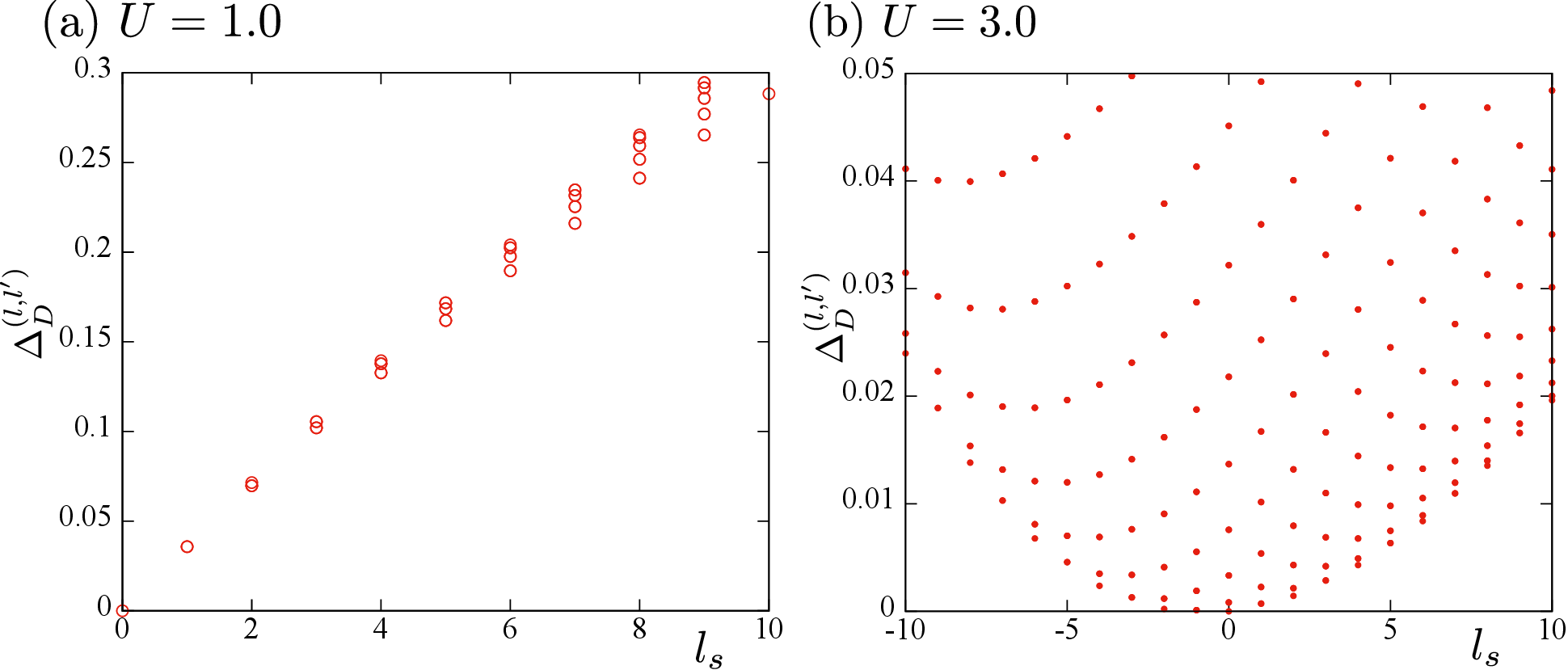}
    \caption{The scaling dimension spectrum with $p=100$ at $\beta=100$ for (a) $U=1.0$ and (b) 3.0. 
    $\Delta_D^{(l,l')}$  is plotted against $l_s\equiv l+l' -2 l^*$ of the horizontal axis.
    The origin of the vertical axis is also sifted by $\Delta_D$.
    In this figure we plot the contributions only form $l, l' \ge 0$, for simplicity. }
    \label{tag_fig5}
\end{figure}

In the metallic phase [Fig.~\ref{tag_fig5}(a)], we have confirmed a linear spectrum, reflecting the linear dispersion of $G^*(i\omega_l)$ with respect to $l$ around $\omega^*\simeq 0$. 
We also find that the low-lying spectrum has a $(l_s+1)$-fold degeneracy within $l, l' \ge 0$. 
As $l_s$ increases, however, such a degeneracy structure gradually collapses, reflecting the deviation of $G^*(i\omega_l)$ from the linear behavior in Fig.~\ref{tag_fig3}.
Note that the region where the linear behavior is observed becomes narrower as $U$ increases within the metallic phase.

In the insulating phase, on the other hand,  $G^*(i\omega_l)$ for $U=3.0$ in Fig.~\ref{tag_fig3} has a round peak at $l^*=19$$(\omega^* = 1.23)$, implying that $l_s$ can be negative.   
Consequently, the spectrum exhibits scattering states of a quadratic envelope with slight asymmetry, as shown in Fig.~\ref{tag_fig5}(b).
Moreover, the overall scale of $\Delta_{D}^{(l,l')}$ in the insulating phase becomes small compared to that in the metallic phase.
We can thus distinguish the Mott transition by the shape of the spectrum structures of scaling dimensions, implying that the bulk phase transition certainly affects the properties of the boundary correlation functions as long as $p$ is finite.

However, we should note that in the $p\to \infty$ limit, the scaling dimension $\Delta_D$ is reduced to the free-electron value, regardless of whether the system is in a metallic or insulating phase.
Also for the excitation, the scale of the dispersive term, $\log |\tilde{t} G^*(i\omega_{l})| /\log p $, is squeezed by $\log p$. 
For the Bethe lattice with a large $p$, the dominant number of electrons is placed in the vicinity of the outer edge boundary, where the distance $x$ in the circumference direction also diverges even for $\bar{n}(\alpha,\alpha')=1$.
 When $p\to \infty$, thus, the scaling dimension associated with the boundary electron may become insensitive to the detailed situation of the deep interior.
As mentioned in Sec. \ref{sec_3}, further studies are needed to clarify the relationship between DMFT and the dimensionality of the corresponding boundary theory.

\section{summary and discussions}

We have explored a holographic renormalization group (RG)-like structure of the dynamical mean-field theory (DMFT) for the semicircle density of states, which can be termed ``AdS/DMFT".
In particular, we recursively determine the branch Green's function of electrons from the outer edge into the interior of the corresponding Bethe lattice network, where the fixed point relation of the recursive RG transformation yields the self-consistent solution of the local Green's function in DMFT.
Moreover, we have introduced an effective AdS$_2$ with the Poincar\'e radius $L=1/\log p$ near the outer edge boundary, by smearing out the lattice nodes in the Bethe lattice.
We have then found that the scaling dimensions for the branch Green's function (or equivalently for the effective chemical potential) and the correlation function of the boundary electron density operators satisfy the scalar field relation (\ref{eq_adsdim}) in AdS$_2$.
This relation of the scaling dimensions is also consistent with $p$-adic AdS/CFT. 
We have then performed DMFT calculations for the Hubbard model with a semicircle density of states;  
 the scaling dimensions corresponding to the metallic and insulating solutions coexist in $2.36 < U < 2.55 $, reflecting that the Mott transition at a finite temperature is of first order. 
We have demonstrated that the scaling dimensions for the excitations detect the features of the metallic and insulating phases, implying that the boundary correlation functions can systematically capture information about the deep interior.
In this paper, we have restricted ourselves to the paramagnetic case. 
However, the Bethe lattice has a bipartite structure, implying that the ground state may have an antiferromagnetic order\cite{Momoi_1998,Chitra_1999,Yanatori_Koga_2016,Chen2025} 
An extension of our RG scheme for the magnetically ordered system is an interesting problem.

In the context of AdS/CFT, the above results suggest that, reflecting the particular nature of the Bethe lattice, DMFT can serve as a prototypical toy model that incorporates quantum fluctuations on the background tree network, which mimics a Euclidian AdS$_2$.
On the other hand, we have not directly discussed the duality with the corresponding boundary CFT, which seems highly nontrivial, since the outer edge nodes of the Bethe lattice exhibit a highly fragmented structure in the circumference direction. 
In this sense, our approach to DMFT should be distinct from the bulk reconstruction based on the boundary CFT\cite{HKLL2006}.
Understanding the nature of the boundary theory corresponding to DMFT is an interesting future problem.

In addition, the holographic RG strucure involved in DMFT provides several interesting implications toward its possible generalizations, which may be relevant to the holography and quantum entanglements.
In the following, we briefly discuss some issues in order.
First, in this work, we have concentrated on the Bethe lattice model, where we could directly introduce the recursive relation of Green's function.
For general lattice models, however, it is not straightforward to write down the recursive relation directly on the real-space lattice network.
In DMFT, instead, the local self-energy is extracted from Green's function, and then the lattice network information is self-consistently implemented into the local Green's function through the Hilbert transformation with respect to the density of states, where the $k$ dependence of the single-particle dispersion is traced out.
The use of the Hilbert transformation suggests that the real-space network structure could be appropriately smeared out without directly referring to the real-space lattice network.
How the DMFT framework generates a curved space coordinate $(r,x)$ relevant to AdS$_2$ would be an important problem.

Next, the effective free-fermionic medium in DMFT is justified for the infinite spatial dimension, which corresponds to $p=(q-1)\to \infty$.
This limit corresponds to $L \to 0$ in AdS$_2$, where the scaling dimension is consistently reduced to the free fermionic value, $\Delta_D=1$.
Since all branches in the Bethe lattice network are equivalent, it does not have an essential metric in the circumference direction.
These observations suggest that the spatial dimension in DMFT is not directly related to the dimension $d$ in AdS$_{d+1}$/CFT$_{d}$.
Although there are $d$-dimensional extensions of boundary theory in the context of $p$-adic AdS/CFT\cite{Gubser2017}, their relevance to DMFT has not yet been clarified.
Another approach to implement spatial dimensions beyond the tree network lattice is to introduce the loop network structure. 
Recently, it has been shown that hyperbolic lattice models, which contain planar loop structures, exhibit several interesting behaviors.\cite{Asaduzzaman2020,Asaduzzaman2022,oku2024,Wang2025,UedaK2007,Krcmar2008}
Moreover, the loop network structure plays an intrinsic role in representing critical quantum fluctuations in the entanglement renormalizaton group and related tensor network algorithms.\cite{MERA2007,TNR2015,LoopTNR2017,GILT2018}
However, their extension to the higher dimensions is still a challenging problem.

Finally, we should note that in DMFT, local electron correlations are properly taken into account at the level of Green's function through the imaginary time path integral.
However, the imaginary-time path integral was done independently of the generation index before mapping to the effective Poincar\'e coordinate (\ref{eq_ads2}). 
Also, the Green's function is a ``classicalized'' quantity, where quantum entanglements at the level of the wavefunction are already averaged out.
These could be a reason why DMFT yielded the relation (\ref{eq_adsdim}), which is equivalent to the Bethe lattice Ising model, although Green's function certainly contains quantum fluctuation effects in the spectrum of the scaling dimensions.
Recently, it has been demonstrated that quantics tensor train and tensor cross-interpolation approaches have successfully captured rich ``entanglement" structures embedded in Green's function of quantum many-body systems~\cite{Shinaoka2023,Ritter2024}.
The theoretical framework of the tensor network has also been developed for classical statistical models~\cite{Oku2023}.
For further improvement of the efficiency of DMFT computations, it may be an interesting problem to modify the network structure in the imaginary time direction from the cylinder shape geometry in Fig.~\ref{tag_fig1}. 
Also, we believe that our holographic RG point of view for DMFT provides a hint for understanding the role of quantum fluctuations in AdS/CFT.

\acknowledgments

One of the authors (K. O.) would like to thank Dr. Y. Inokuma for valuable discussions on dynamical mean-field theory.
This work is partially supported by Grant-in-Aid for Transformative Research Areas titled  ``Natural Laws of Extreme Universe" (KAKENHI Grants, Nos. JP21H05182 and JP21H05191), "Foundation of Machine Learning Physics" (No. JP25H01521), and "Hyperuniform Aperiodic Materials" (No. JP25H01398) from JSPS of Japan. 
It is also supported by JSPS (KAKENHI Grant, No. JP22K03525) and JST-CREST (No. JPMJCR24I1).
The numerical simulations have been performed using some of the ALPS libraries~\cite{alps}.

\bibliography{dmft}

\appendix

\section{free electrons}
\label{appendix_A}

As well established for the free-electron system on the Bethe lattice\cite{DMFT_RMP1996}, we can explicitly solve the recursive relation for the branch Green's function, which is instructive for understanding a holographic structure in DMFT. 

\subsection{fixed point Green's function}
 Assuming all branches are equivalent and omitting the branch index $\alpha$, we can exactly represent the recursive relation as 
\begin{align}
        G_{n-1 \sigma}(z) & = \frac{1}{  z + \mu  - pt^2  G_{n\sigma}(z)} \, ,
        \label{eq_freeRG}
\end{align}
where $\mu$ denotes a chemical potential, 
Then, the self-consistent solution is straightforwardly obtained as
\begin{align}
    {G}^*_\sigma(z) = \frac{1}{2pt^2}\left( z+ \mu \mp \sqrt{(z+\mu)^2 -4pt^2} \right)\, .
\label{eq_betheG_solution}
\end{align}
Note that the ``$-$'' branch should be adopted for $\mathrm{Im}(z) >0$ while the ``+" branch for $\mathrm{Im}(z)<0$, because of the analyticity in the $|z|\to \infty $ limit. 
In addition, we should note that ${G}^*_\sigma(z)$ has the branch cut for $|z| \le \sqrt{4pt^2}$ on the real axis.

Here, recall that $q$ numbers of descendant branches meet ($p=q-1$) at the center node of $n=0$, as in Eq. (\ref{eq_def_Z}).
Thus, Green's function at $n=0$ should be recovered from Eq.(\ref{eq_betheG_solution}) as
\begin{align}
    {G}^\mathrm{o}_{\sigma}(z) 
    = \frac{(q-2)(z+\mu) \mp q\sqrt{(z+\mu)^2 - 4 p t^2}}{2(q^2t^2 -(z+\mu)^2)} \, ,
    \label{eq_bulk_dos}
\end{align}
which corresponds to the bulk Green's function for the Bethe lattice model. 
Here, note that we took the $-$ branch for Eq. (\ref{eq_betheG_solution}).
The bulk density of states $\rho(\omega)$ extracted from Eq. (\ref{eq_bulk_dos}) is
\begin{align}
    \rho(\omega) = -\frac{1}{\pi}\mathrm{Im} {G}^\mathrm{o}_\sigma(\omega + i 0) = \frac{q\sqrt{4pt^2 - (\omega+\mu)^2}}{2\pi (q^2t^2 -(\omega+\mu)^2)} \, ,
\end{align}
which is defined in the region $-2\sqrt{p} t - \mu <\omega < 2\sqrt{p} t -\mu$.
Taking the $p,q\to \infty$ limit with scaling $\tilde{t}= t/\sqrt{q}$, we can then recover the semicircle density of states for the Bethe-lattice model,
\begin{align}
     \rho(\omega) = \frac{1}{2\pi \tilde{t}^2} \sqrt{4\tilde{t}^2 - (\omega+\mu)^2} \,
     \label{eq_dos_appdx}
\end{align}
for $-2\tilde{t}-\mu <\omega  < 2\tilde{t} -\mu $.
If we introduce the half bandwidth, $\tilde{D}\equiv 2 \tilde{t}$,   Eq. (\ref{eq_dos_appdx}) becomes $\rho(\omega)$ used for DMFT computations in Sec. \ref{sec_numerical} in the main text.

\subsection{Convergence to the fixed point}

In general, it may be nontrivial whether the recursion relation (\ref{eq_k-effectivemedium}) and (\ref{eq_bethe_Gimpurity}) converges to the self-consistent solution. 
For the free-electron system, however, we can precisely analyze a perturbation around the fixed-point solution.
This analysis also provides an intuitive understanding of the holographic RG-like structure for the Bethe lattice model.
Here, we consider how the deviation of ${G}^*_{n\sigma}(z)$ propagates through the recursion relation.
Assuming no $\alpha$ dependence,  we may write
\begin{align}
      {G}_{n\sigma}(z) = {G}^*_\sigma(z) +\delta {G}_{n\sigma}(z)\, .
\end{align}
Lenarlizing Eq. (\ref{eq_freeRG}) with respect to $\delta{G}$, we then obtain
\begin{align}
    \delta {G}_{n-1\sigma\alpha}(z) 
    & = \lambda(z)\delta G_{n\sigma}(z) \, ,
\end{align}
where
\begin{align}
    \lambda(z) =\frac{z+\mu - \sqrt{(z+\mu)^2 - 4pt^2}}{z+\mu + \sqrt{(z+\mu)^2 - 4pt^2}} \, .
    \label{eq_lambda_z}
\end{align}
If $ -2 \sqrt{p}t < z+\mu < 2\sqrt{p} t $ on the real axis, we thus have $|\lambda|=1$, where ${G}^*_\sigma(z)$ is not analytic.
Away from the real axis, we have $|\lambda| < 1 $, which implies that the recursive relation (\ref{eq_freeRG}) has a stable fixed point.
In particular, assuming a free electron at the outer edge node($n=N$),  we have 
\begin{align}
    {G}_{N\sigma}(z) = \frac{1}{z+\mu} \, .
\end{align}
Then we have
\begin{align}
    \delta {G}_{N\sigma}(z) &=  {G}_{N\sigma}(z) - {G}^*_{\sigma}(z) 
    =   -\frac{1}{z+\mu} \lambda(z) \, ,
\end{align}
which leads us to ${G}_{n\sigma}(z) = - \frac{\lambda (z)^{N-n+1}}{z+\mu} $.

In order to see the convergence of $G_{n\sigma}$ in a general situation, it is helpful to consider the mathematical structure of the recursive relation (\ref{eq_freeRG}).
For simplicity, we assume that $t=1$ and all branches are equivalent.
We also rewrite $z + \mu \to z$.
Then, the intrinsic part of the recursive relation can be described by conformal mapping, 
\begin{align}
    \zeta'= f_z(\zeta) \equiv \frac{1}{z-p \zeta } \, ,
\end{align}
where $\zeta$ and $\zeta'$ correspond to ${G}_{n\sigma}$ and ${G}_{n-1\sigma}$, respectively.
An essential point is that $f_z(\zeta)$ is a special case of the M\"obius transformation:
\begin{align}
    f_z(\zeta) = \frac{a\zeta+b}{c\zeta + d}
\end{align}
with $a=0$, $b=1$, $c=-p$ and $d=z$.
Thus, the recursive mapping for the parameters can be represented in the matrix form
\begin{align}
    f_z \circ f_z \circ \cdots f_z = A A \cdots A
\end{align}
with
\begin{align}
A \equiv
\begin{pmatrix}
0 & 1 \\
-p& z
\end{pmatrix} \, .
\end{align}
Note that $ad- bc = p$.
The eigenvalue of $A$ is given by $\Lambda_\pm(z) = \frac{1}{2}(z \pm \sqrt{z^2-4p})$, which is consistent with the fixed point of Eq. (\ref{eq_betheG_solution}).

Moreover, the M\"obius mapping rescales the radius of any circles around the fixed point by 
\begin{align}
|\lambda(z)|=|\frac{\Lambda_-}{\Lambda_+}| = |\frac{z - \sqrt{z^2-4p}}{z +  \sqrt{z^2-4p}}|\, .
\end{align}
Thus, an arbitrary initial Green's function except for $G_{N\sigma} = z/p$ (singular point of $f_z$) can converge to the fixed-point solution of Eq. (\ref{eq_betheG_solution}) exponentially, if $z$ is not in $-2\sqrt{p}< z<2\sqrt{p} $ on the real axis. 
Here, note that the most relevant (slowest convergence) mode is given by $\lambda(\omega_l)$ with $l=0$ or $-1$ as discussed for Eq. (\ref{eq_corr_asympt}).

In DMFT, the free fermionic factorization of many-body Green's functions is assumed.
As in Fig.~\ref{tag_fig3}, the resulting fixed point solution enjoys $|\lambda(z)|<1$ for $\mathrm{Im}(z)\ne 0$.
This is an empirical reason why the DMFT iteration has a stable convergence to the bulk fixed-point solution even for $U\ne 0$.

\end{document}